\documentclass[11pt]{elsart}

\usepackage{graphicx}
\usepackage{epsf}
\usepackage{wrapfig}
\usepackage{here}

\begin{document}

\begin{frontmatter}

\title
{Performance of the Stereoscopic System of the \\ 
HEGRA Imaging Air \v{C}erenkov Telescopes:\\
Monte Carlo Simulations \& Observations}

\author[1]{A. Konopelko},
\author[1]{M. Hemberger},
\author[1]{F. Aharonian},
\author[1]{A. Daum}, 
\author[1]{W. Hofmann}, 
\author[1]{C. K\"ohler},  
\author[1]{H. Krawczynski},  
\author[1]{H.J. V\"olk}, 
\author[2]{A. Akhperjanian},  
\author[3]{J. Barrio\thanksref{4}},  
\author[1]{K. Bernl\"{o}hr},
\author[6]{H. Bojahr},
\author[4]{J. Contreras}, 
\author[4]{J. Cortina}, 
\author[5]{T. Deckers}, 
\author[3]{S. Denninghoff},
\author[3]{J. Fernandez\thanksref{4}},  
\author[4]{V. Fonseca}, 
\author[4]{J. Gonzalez}, 
\author[7]{V. Haustein}, 
\author[7]{G. Heinzelmann}, 
\author[1]{G. Hermann}, 
\author[1]{M. He\ss}, 
\author[1]{A. Heusler}, 
\author[6]{H. Hohl},
\author[3]{I. Holl}, 
\author[7]{D. Horns},  
\author[1]{R. Kankanian\thanksref{2}},  
\author[6]{M. Kestel},
\author[1]{J. Kettler},
\author[5]{O. Kirstein}, 
\author[3]{H. Kornmayer},  
\author[3]{D. Kranich}, 
\author[1]{H. Lampeitl}, 
\author[7]{A. Lindner}, 
\author[3]{E. Lorenz}, 
\author[6]{N. Magnussen},  
\author[6]{H. Meyer}, 
\author[3]{R. Mirzoyan\thanksref{4}},  
\author[6]{H. M\"oller}, 
\author[4]{A. Moralejo}, 
\author[4]{L. Padilla}, 
\author[1]{M. Panter}, 
\author[3]{D. Petry\thanksref{6}}, 
\author[3]{R. Plaga}, 
\author[1]{A. Plyasheshnikov},
\author[7]{J. Prahl}, 
\author[3]{C. Prosch}, 
\author[1]{G. P\"uhlhofer}, 
\author[5]{G. Rauterberg}, 
\author[6]{W. Rhode}, 
\author[7]{A. R\"ohring}, 
\author[2]{V. Sahakian}, 
\author[5]{M. Samorski}, 
\author[4]{J. Sanchez}, 
\author[7]{D. Schmele},
\author[6]{F. Schroeder},
\author[5]{W. Stamm}, 
\author[6]{B. Wiebel-Sooth}, 
\author[1]{C. A. Wiedner}, 
\author[5]{M. Willmer}, 
\author[1]{H. Wirth},
\author{(HEGRA collaboration)}

\bigskip

\address[1]
{Max-Planck-Institut f\"ur Kernphysik, D-69029 
Heidelberg, Germany}
\address[2]
{Yerevan Physics Institute, 375036 Yerevan, Armenia}
\address[3]
{Max-Planck-Institut f\"ur Physik, F\"ohringer Ring 6, D-80805
Munich, Germany}
\address[4]
{Facultad de Ciencias Fisicas, Universidad Complutense,
E-28040 Madrid, Spain}
\address[5]
{Universit\"at Kiel, Inst. f\"ur Kernphysik, D-42118 Kiel, Germany}
\address[6]
{BUGH Wuppertal, Fachbereich Physik, D-42119 Wuppertal, Germany}
\address[7]
{Universit\"at Hamburg, II. Inst. f\"ur Experimentalphysik, 
D-22761 Hamburg, Germany}

-----------------------------------------

\pagebreak

\begin{abstract}

Based on the Monte Carlo simulations we have studied the performance 
of the HEGRA system of imaging air \v{C}erenkov telescopes (IACTs) in its present 
configuration of 4 IACTs as well as in its future final configuration of 5 IACTs. 
Here we present the results on the basic characteristics of the IACT system 
which are used in the standard data analysis procedure, i.e., the collection areas, 
the detection rates, the angular resolution, the energy resolution, and the 
$\gamma$/hadron-separation efficiency. By comparing several key Monte Carlo predictions 
with experimental results it is possible to check the accuracy of the simulations. The 
Monte Carlo results concerning hadron-nuclear showers are tested with the recorded cosmic 
ray events and the results concerning photon-induced showers are tested with a large data 
sample of $\gamma$-rays observed from BL Lac object Mkn 501 during its high flaring 
activity in 1997. Summarizing the simulations and current observations we give the basic 
recommendations of using the instrument and the major values of its sensitivity. 

\noindent
{\it PACS:} 95.55.Ka; 95.55.Vj; 96.40.Pq

\noindent
{\it Keywords:} Imaging Air \v{C}erenkov Technique; Very High Energy Gamma Ray Astronomy; 

\end{abstract}
\end{frontmatter}

\section{Introduction}

The HEGRA collaboration is close to completing an array of five imaging air \v{C}erenkov 
telescopes (IACTs) located on the Roque de los Muchachos, Canary Island La Palma 
($28.8^{\circ} \, \, \rm N, 17.9^{\circ}$). The telescope array, primarily designed for 
stereoscopic observations of the $\gamma$-radiation at energies above several hundred GeV 
[1] is formed by five identical IACTs - one at the center, and four others at the corners 
of a 100~m by 100~m square area. The multi-mirror reflector of each system telescope has 
an area of $8.5 \, \rm m^2$. Thirty 60~cm diameter front aluminized and quartz coated 
spherical mirrors with focal length of about 5~m are independently installed on an almost 
spherical frame of an alt-azimuth mount. The FWHM of the point spread function of the 
reflector is better than 10 arcminutes. Each telescope is equipped with a 271-channel 
camera [2] with a pixel size of about $0.25^{\circ}$ which results in the telescope's effective 
field of view of $\simeq 4.3^{\circ}$. The digitization of the PMT pulses is performed 
with a 120 MHz FADC system. The system trigger demands at least two neighboring pixels above 
the threshold in each of at least two telescopes. The present system of four telescopes is 
taking data in the stereoscopic mode since 1996 [3]. The whole telescope array of 5 IACTs 
is planned to run as a single detector at the end of 1998.

The basic concept of the HEGRA IACT array is the stereoscopic approach based on the 
simultaneous detection of air showers by $\geq 2$ telescopes, which allows precise 
reconstruction of the shower parameters on an event-by-event basis, superior rejection of 
hadronic showers, and effective suppression of the background light of different origins 
(night sky background, local muons, {\it etc.}). The recent observations of the Crab Nebula 
[3] and Mkn~501 [4] by the IACT system strongly support the theoretical expectations concerning 
the features of the stereoscopic imaging technique [5,6].

Due to the lack of a calibrated very high energy $\gamma$-ray beam, detailed 
Monte Carlo simulations are usually used for the design studies as well as for 
performance studies of the imaging atmospheric \v{C}erenkov experiments. For example, 
new data analysis methods are developed and are tested with Monte Carlo 
simulations before being applied to real data. The measurement of absolute $\gamma$-ray flux 
and energy spectra of the established $\gamma$-ray sources as well as the determination of 
upper limits for the quiet objects heavily rely on Monte Carlo predictions of the detector  
performance. 
In the past, the comparison of the characteristics of recorded cosmic ray (CR) 
events with the characteristics of the Monte Carlo simulated hadron-induced 
air showers used to be the most reliable way to test the predictive power 
of the Monte Carlo simulations (e.g., [7]). Once the telescope response to CR-induced air 
showers is well understood, the Monte Carlo predictions for the $\gamma$-ray- 
induced shower can be performed with a high degree of confidence. 
The situation changed dramatically with the observation of the high flaring 
activity of Mkn 501 in 1997. The observations with the HEGRA system of 4 IACTs 
($\simeq 110$ h) provided a large data base of  $\gamma$-rays ($\geq$ 30000) 
with unprecedented signal-to-noise ratio. Several observational key characteristics 
of $\gamma$-ray- induced air showers can be measured with small statistical 
uncertainties. The agreement of these 
key characteristics of $\gamma$-ray induced air showers in data and Monte 
Carlo simulations substantially strengthened the reliability of the simulations.  

In this paper we discuss the standard data analysis procedure and the results of 
the detailed Monte Carlo simulations 
for the currently operating system of 4 IACTs as well as for the complete 
HEGRA array of 5 IACTs. Special attention has been paid on the proper 
simulation of the camera and electronics of the \v{C}erenkov telescopes (see Section 2). 
Detailed comparisons of the detected cosmic ray and $\gamma$-ray-induced air 
showers with simulations have been made to understand the performance of 
the detector (Section 3). The basic 
characteristics of the HEGRA IACT 
system were calculated (Section 4). Finally, we discuss the resulting sensitivity 
of the HEGRA IACT array with respect to TeV $\gamma$-rays (Section 5) and summarize 
the basic recommendations for the use of the instrument.  

\section{Simulations}

In the present calculations the complete Monte Carlo simulation procedure 
was divided into two steps. In the first step an extended library of 
simulated air showers, induced by different primaries, is produced. 
In the second step simulation of the response of the 
telescope camera pixels is applied to all generated events. The most time 
consuming step is evidently the first one, whereas the second step is 
relatively fast. This division allows to apply the detector simulation 
procedure several times to the generated showers in order to tune the Monte 
Carlo simulations to the hardware status of the telescopes (e.g. 
trigger threshold, mirror adjustment etc.) Here we discuss the 
major features of this two-step simulation procedure.

\subsection{Shower Simulation}

The generation of the air showers has mainly been performed using 
the ALTAI Monte Carlo code [7]. For the 
electromagnetic cascade this code has implemented an algorithm 
based on the analytical probability distributions of the electron (positron) 
transport in the multiple-scattering segments. This algorithm substantially 
reduces the computational time needed for simulation of a single shower. 
The proton-nuclei cascade in the atmosphere is simulated according to the 
radial scaling model (RSM) based on accelerator data [8]. The air 
showers induced by the primary nuclei are simulated in a model of independent 
nucleon interactions for the fragmentation at the projectile-nucleus. The 
fragmentation of the colliding nuclei is processed according to the 
probabilities of different fragmentation channels. 
We studied the influence of the proton-nuclei cascade model on the  
observable shower parameters using the additional simulations with the 
CORSIKA code [9] which has implemented the HDPM model 
for the simulation of the nucleus-nucleus interactions.

The shower simulation is carried out at the level of single \v{C}erenkov 
photons. A certain fraction ($k \simeq 0.2$) of the \v{C}erenkov photons of a 
shower which hit the telescope reflector are stored with full information, 
i.e. the arrival time, the arrival direction, and the impact coordinates in 
the reflector frame. By this means it is possible to apply the complete 
detector simulation procedure to all showers which have been processed 
in this way. 
The Monte Carlo library contains air showers induced by the primary 
$\gamma$-rays, protons, helium and other nuclei belonging to CNO, heavy and 
very heavy nuclei groups. The primary energy of the showers is randomly 
distributed inside 14 fixed energy bins covering the energy range from 100 
GeV to 100 TeV. The events are used with weights according to some chosen 
primary spectra. The simulations have been performed for zenith angles 
$0^\circ, \, 30^\circ, \, 45^\circ$. For each type of primary particle and 
for each zenith angle approximately $10^5$ showers were simulated. 
The actual setup of HEGRA IACT telescopes [3] was used in the simulations. 
The position of the shower axis in the observation plane was 
uniformly randomized over the area limited by the radius $R_0$ with respect 
to the central telescope. The radius $R_0$ was chosen between 250 and 450 m, 
increasing with shower energy and inclination angle. For the CR air showers 
the additional randomization over the solid angle around the telescope axis 
with the half opening angle of $3.5^\circ$ has been introduced in order to reproduce 
the isotropic distribution of the CR events over the camera field of view.

\subsection{Detector Simulation}

The detector simulation procedure accounts for all 
efficiencies involved in the process of the \v{C}erenkov light propagation which 
starts with emission of a \v{C}erenkov photon and ends with the digitization of 
the PMT signal (see for details [10]). The list of the effects 
which are important in this respect contains: (i) mirror reflectivity, 
modelled with the raytracing technique or in phenomenological way using 
the measured functions of the light spot distortion in the camera focal 
plane; (ii) the light absorption in the plexiglass panel covering the 
camera; (iii) the acceptance of the funnels placed in front of the 
photomultipliers (PMTs); (iv) the photon-to-photoelectron 
conversion inside the PMTs (EMI 9083R) taking into account a measured  
single photoelectron spectrum. The overall efficiency of the 
photon-to-photoelectron conversion is of $\sim 0.1$. 
By analogy with the experiment the structure of the readout based on the 
120 MHz FADC data acquisition and the multiple-telescope trigger scheme 
[11] were implemented in the Monte Carlo simulations. Note, 
that this procedure takes into account the arrival times of the \v{C}erenkov 
light photons hitting the telescope reflector. The basic characteristics of 
the performance of the telescope hardware, for instance the number of camera 
pixels read out for the triggered and none triggered telescopes, 
the single pixel rate and the single pixel trigger rate, the ratio between first 
and second maximum pixel amplitudes in the image etc have been directly 
compared between Monte Carlo and data [10]. 

\section{Parameters of the Cosmic Ray Air Showers} 

The detection rate of the IACT system is mainly determined by the 
isotropic flux of primary cosmic ray protons and nuclei. The system 
is triggered if the shower produces a sufficiently high number of 
\v{C}erenkov photons to trigger at least two telescopes. Since the 
number of \v{C}erenkov photons produced by a shower is to first order 
approximation proportional to the energy of the primary particle, 
the trigger condition determines the energy threshold of the IACT 
system. At larger shower axis distances the \v{C}erenkov photon 
density decreases rapidly. With increasing energy of the primary 
particles more and more showers are able to fulfill the trigger  
criteria, although they have impact distances far away from the 
telescope system. However, due to the steep primary energy spectrum 
of cosmic rays $dJ_{cr}/dE \propto E^{-2.75}$ the contribution of the 
high energy showers ($\rm E>20\, TeV$) to the total cosmic ray detection 
rate is rather small even though they are collected over a larger area. 

\subsection{Cosmic Ray Detection Rate}

The basic characteristics of a single \v{C}erenkov telescope at hardware level 
can be calculated using the procedure described in [12]. 
For a system of IACTs the only difference stems from the performance of the 
multi-telescope trigger. For this purpose, first the local trigger condition 
at each telescope as $\rm 2nn/271>q_0$ (the signals in at least two neighboring 
pixels exceed $q_0$) has to be fulfiled. The trigger threshold $q_0$ is measured in 
photoelectrons. The system trigger demands for each 
individual event coincident trigger signals from at least 
$N$ telescopes (N/5, N = 2, 3, 4, 5). 
Assuming the energy spectrum and the chemical composition of the primary 
cosmic rays [13], the cosmic ray detection rate can be computed with the following 
formula:
\begin{equation}
R_{cr} = \int^{\Omega_0}_0 d \Omega\,\, 2 \pi \int^{R_0}_0 RdR \int^{E_0}_0 
\frac {dJ_{cr}} {dE d\Omega dS} P_{cr}(R,E,\theta)
\end{equation} 
where $E$ is the primary energy of a cosmic ray, $R$ is the impact distance 
of the shower induced by the cosmic ray, $\theta$ is the angle of the 
shower axis with respect to the telescope axis, and 
$P_{cr}(R,E,\theta)$ is the probability of the shower to trigger the 
telescope system.  

The performance of the multi-telescope trigger of the HEGRA IACT system was 
discussed in detail in [11]. In Table 1 we show the 
hardware detection rates for the different trigger thresholds as measured 
with the HEGRA telescope system [11] together with the 
results of the Monte Carlo simulations. The measured and computed rates 
are in a good agreement which confirms that the trigger procedure has been 
modelled correctly. Note that the absolute accuracy of the measured rate is about 
0.2 Hz. The Monte Carlo predicted rates strongly rely on the used chemical composition 
and fluxes of the cosmic rays. So the accuracy of the Monte Carlo predicted 
rates is better than 20 \%. 
The calculations of CR detection rates have been done also for the complete HEGRA 
system of 5 IACTs (see Table 2). One can see that the hardware detection rate for 
the complete array is expected to be approximately $\simeq 1.4$ times larger than 
for the currently operating 4 IACT system.

\subsection{Cosmic Ray Shower Images}

The standard method to parametrize the \v{C}erenkov light images was primarily 
introduced in [14]. It is based on the second moment analysis of 
the two-dimensional angular distributions of \v{C}erenkov light flashes, 
sampled with pixels (PMTs) of finite solid angular extension [15].
The effective technique (supercuts) to extract the shower image from the 
measured  matrix of the pixels amplitude was suggested in [16].
This method provides an effective $\gamma$/hadron separation 
and has been extensively used by several single \v{C}erenkov telescopes around 
the world. For the system of IACTs the performance of the method can be  
substantially improved, as will be shown below.

The distributions of the second-moment image parameters depend crucially on 
the amount of background light per pixel. A slight overestimate or 
underestimate of the background light content dramatically change the 
distributions. The detailed detector simulation procedure is provided to account 
the exact background content in the camera pixels. 						  
In Figure 1 we show the distributions of the second-moment parameters of the 
cosmic ray images measured in the OFF region of Mrk 501 observations (dark 
sky region) by the HEGRA system of 4 IACT telescopes and the Monte Carlo 
simulations.  It can be seen in Figure 1 that the Monte Carlo 
simulated images fit the measured images quite well. 

In general, the distributions of the shape parameters of the \v{C}erenkov light 
images depend on the model of the development of the proton-nuclei cascade in 
the atmosphere and also on the model of the nucleus-nucleus interaction. To study 
this effect the series of simulations have been performed using the ALTAI 
code [7] and CORSIKA code (with HDPM model) [8] 
in the fragmentation model as well as under assumption of a simple 
superpositon model of nucleus-nucleus interactions. 
The results show a good agreement of two simulation codes, despite of the 
very different models which are used for the simulations of the proton-
nuclei component of the air showers.
The distributions shown in Figure 1 have been produced assuming a certain 
chemical composition of the primary cosmic ray [13]. 

\section{Imaging of the Gamma-Ray Air Showers}

The hardware detection rate of the cosmic ray air showers dominates by at 
least two orders of magnitude over the detection rate of the $\gamma$-rays. 
To extract the $\gamma$-ray signal at sufficient confidence level, 
a special analysis is used to 
suppress significantly the rate of the cosmic ray events. This analysis 
is based on the application of several software cuts related to the orientation 
and the shape of the \v{C}erenkov light images. 
Assuming an integral flux and an energy spectrum index of a $\gamma$-ray 
source, the detection rates of the $\gamma$-ray-induced air showers before and 
after application of the software cuts can be calculated.  

\subsection{Collection Areas and Detection Rates}

One of the major advantages of the ground based \v{C}erenkov technique 
in comparison with satellite observations is that the VHE 
$\gamma$-ray-induced air showers can be detected at large distances 
($\geq 100$ m) of the shower core from the telescope. That yields a 
high detection rate of the $\gamma$-ray-induced air showers which are 
distributed over the large area of $S_\gamma \simeq 10^9 \, \rm cm^2$  
around the telescope site. The collection area for the $\gamma$-ray-induced 
air showers is calculated as 
\begin{equation}
S_\gamma(E) = 2 \pi \int_0^\infty P_\gamma(E,r)rdr
\end{equation}
where $P_\gamma(E,r)$ is the trigger efficiency of $\gamma$-ray-induced 
air showers of primary energy $E$ and impact distance $r$. 
The collection area $S_\gamma(E)$ for showers of primary energy E,  
is mainly determined by the effective area of the telescope reflector 
$S_{ph.e.} = S_m \cdot \chi_{ph.e.}$, where $S_m$ is the total mirror area and 
$\chi_{ph.e.}$ is the efficiency of the photon-to-photoelectron conversion of 
the camera channels. Given a fixed mirror area, a maximum collection area is   
achieved by reducing the trigger threshold of the telescopes. The latter is 
limited at the lower end by the fluctuations of the background light in each 
camera pixel. 

In Figure 2 the collection areas for the complete HEGRA system of IACTs are 
shown for the conventional trigger criteria. The strong increase of 
collection area in the energy range $\leq 1\rm \, TeV$ changes to a 
logarithmic growth at higher energy. The behaviour of the collection areas is 
determined by the shape of the lateral distribution of the \v{C}erenkov light 
pool at the observation level. The density of \v{C}erenkov light photons at 
plateau for impact distances up to 125 m (for air showers observed at angles 
close to zenith) is roughly proportional to the primary shower energy whereas 
beyond 125 m the density of the \v{C}erenkov light density decreases rapidly. 
In observations at large zenith angles ($\ge 30$ degree) the collection area 
decreases at low energies ($E\le \rm \, 3 \,TeV$) but increases at higher 
energies (E$>$5~TeV). The reason for that is that air showers at the larger 
zenith angles develop higher in the atmosphere. This is clue to the fact that the 
same amount of the \v{C}erenkov light (neglecting the increase in absorption) is 
produced by a shower, but is scattered over the larger area at the 
observation level which provides decreasing the density of \v{C}erenkov photons. 
Thus low energy air showers at large zenith angles cannot trigger the telescopes 
but at the same time the high energy air showers have much larger collection 
area due to the large size of the \v{C}erenkov light pool at observation level. 

The HEGRA system of 5 imaging air \v{C}erenkov telescopes has been designed for  
effective observation of $\gamma$-rays with the primary energy of several 
hundred GeV in stereoscopic mode with telescopes of relatively small 
mirror area, 8.5 $\rm m^2$. The system trigger based on the simultaneous 
detection of the shower images in several telescopes (at least 2) forces down 
the trigger threshold and consequently the energy threshold of the IACT 
system. Usually, the energy threshold is determined as the energy at 
which the detection rate of observed $\gamma$-ray showers reaches its maximum. 
For convenience we use in the following rate calculations a 
$\gamma$-ray spectrum according to:  
\begin{equation}
dJ_\gamma/dE = A\cdot E^{-\alpha_\gamma},\,
J_\gamma(>1\, \rm TeV) = 10^{-11}\, cm^{-2}s^{-1}.
\end{equation}
For a certain spectrum index $\alpha_\gamma$ the $\gamma$-ray detection rate 
is calculated as
\begin{equation}
R_\gamma = \int_0^\infty (\frac{dR_\gamma}{dE}) dE = 
\int_0^\infty (\frac{dJ_\gamma}{dE}) S_\gamma(E) dE
\end{equation}
where $(dR_\gamma/dE)$ [$\rm Hz \, TeV^{-1}$] is the differential 
$\gamma$-ray detection rate.

Under the assumption of a differential index, $\alpha_\gamma$, of the $\gamma$-ray energy 
spectrum one can calculate the differential $\gamma$-ray detection rate $(dR_\gamma/dE)$ 
(see Figure 3). The peak of the differential detection rate slightly shifts to higher 
energies with increase of the trigger multiplicity because low energy events cannot 
effectively trigger several telescopes. The integral detection rates of the $\gamma$-ray-induced 
air showers for different trigger multiplicities are presented in Table 3. Note that in the 
case of a steep $\gamma$-ray spectrum (e.g. $\alpha_\gamma$ = 3.0) for the system 
trigger 2/5 the $\gamma$-ray detection rate is much higher as compared with the operation 
for the 3/5 trigger mode.  
It is seen from Figure 3 that 
for the trigger conditions 2/5, 2nn/271$>q_0$ ph.e. the 
differential detection rate peaks at an energy of $\simeq 500 \rm \, GeV$ which 
is identified as the energy threshold of the instrument. 
The energy threshold for observations at large zenith 
angles significantly increases (see Figure 4). At the same time, the rate of  
high energy events detected at large zenith angles could even exceed the corresponding rate 
in observations at the nominal zenith angles due to the significant increase of the  
collection area with increasing zenith angle. 

Our Monte Carlo studies show that $\gamma$-ray air showers detected at large impact 
distance from the telescopes cause some difficulties for a reliable selection of $\gamma$-ray 
events. For plane-parallel $\gamma$-ray flux the \v{C}erenkov light image shifts to the 
camera edge for large impact distances. These images are partially cut by the 
camera edge and cannot be used for an accurate reconstruction of the shower parameters 
(orientation of the shower axis in space, shower core location etc.). 
Thus for better evaluation of the energy spectrum it is useful to set a restriction on the 
reconstructed impact radius 
from the center of the system. This restriction influences mainly the collection areas at 
high energies. Above a certain energy the effective collection area is then determined 
simply by the geometrical area around the IACT system. The restriction on the impact 
distance for $r<200$ m in the case of steep spectrum does not significantly change the 
detection rate. For the case of a flat energy spectrum, especially at observations 
at large zenith angles, it is an advantage to collect $\gamma$-ray-induced air 
showers at large distances to the center of the IACT system.  

\subsection{Reconstruction of Shower Arrival Direction}

The simultaneous observations of the air showers with $\geq 2$ imaging air \v{C}erenkov 
telescopes offers the possibility to reconstruct the orientation of the shower axis with 
respect to the telescope axis [18]. The general approach is based 
on the superposition of the several images in one common focal plane in order to derive the 
intersection point of the major axis of the ellipsoid-like images. This intersection 
point determines the shower direction [6]. If the \v{C}erenkov 
telescopes are directed towards the object, the reconstructed source position in the camera 
field of view for the $\gamma$-ray-induced air showers has to be in the center of the camera 
focal plane. The currently operating HEGRA system of IACTs performs so-called {\it wobble mode} 
observations. The position of the source in the camera focal plane is offset by $0.5^\circ$  
from the camera center (on declination) and consequently rotates depending on the azimuth 
angle. This approach gives the possibility to perform continuous ON source observations, 
whereas the OFF region can be chosen in $1^\circ$ offset from the source position 
[3]. In present simulations the \v{C}erenkov light images were shifted by 
$0.5^\circ$ from the center of the focal plane with the correlated randomization over the 
azimuth. 

The difference between the true and reconstructed position of the $\gamma$-ray source 
in the camera field of view, $\Theta$, is a measure of the angular resolution 
of the system of IACTs. The distributions of $\Theta^2$, both for the simulated and observed  
$\gamma$-ray showers from Mrk 501 are shown in Figure 5. One can see from Figure 5 
both distributions match and both show a prominent peak around the source 
position. Our Monte Carlo studies   
show that the tail of the distribution at large $\Theta^2$ is due to the air 
showers with the core positions close 
to the line connecting two or three telescopes. In such a case, the images in the telescopes 
are almost parallel to each other and the reconstruction procedure leads to significant error 
in the evaluated shower direction because of a small intersection angle. Note that in  
the reconstruction procedure we require at least 3 telescopes have to be triggered (in addition 
we require also a minimum number of photoelectrons in the image of 40 ph.e.). 

The angular resolution of the system of IACTs can be characterized quantitatively 
by the acceptance of the $\gamma$-ray-induced air showers, $\kappa^{dir}_\gamma$, after the 
application of the fixed angular cut on $\Theta^2$. In Table 4 the data on the acceptance, 
$\kappa^{dir}_\gamma$, of the $\gamma$-rays for three angular cuts $\Theta^2\leq 0.03, \, 0.05$ 
and 0.1 $\rm [deg^2]$ are shown for the simulations at different zenith angles 0, 30, 45 
degree. In general, for observations at large zenith angles the shower is far from the observer 
and the \v{C}erenkov light images detected at large zenith angles have a smaller angular size, 
they are closer to the camera center and show almost circular shape. These changes in the 
image topology lead consequently to larger errors in the reconstruction of the shower direction.
The comparison of the angular resolution for two different trigger multiplicities, 2/5 and 
3/5, show some increase in the $\gamma$-ray acceptance applying the same angular cuts for 
the higher trigger multiplicity -- 3/5. However, because of the energy dependence of the angular 
resolution two different system triggers have to be considered as complementary in 
observations of $\gamma$-ray sources with very different spectral features.

The angular resolution noticeably depends on the impact distance of the shower core from the center to 
the IACT system. For impact distances $r\leq 125$ m the angular 
resolution slightly improves with increasing the energy of the $\gamma$-ray showers, because the 
images on average contain more light and the image orientation is better determined.
The angular resolution, $\delta \Theta$ (one standard deviation of the Gaussian distribution 
on $\theta$), for $\sim$1 TeV $\gamma$-ray showers is of $0.11^\circ$ and $0.09^\circ$ at $\sim20$ 
and 200 m, respectively. Beyond $\sim 120$ m 
the angular resolution decreases at higher energies. This can be explained by the 
distortion of the images by the edge of the limited camera field of view. The impact distances 
for high energy $\gamma$-rays correspond to large shifts of the images from the 
center of the camera focal plane (the high energy air showers occur deeper in the 
atmosphere). For $\sim10$ TeV $\gamma$-ray showers detected at impact distances of 200 m the 
angular resolution is $0.14^\circ$. 
 
The advanced angular resolution of the HEGRA system of IACTs is a very effective tool for 
suppression of the isotropic cosmic ray background. The acceptances of  
cosmic ray air showers after application of an angular cut, $\Theta^2<\Theta^2_0$, 
evaluated from the data taken with the currently operating HEGRA system of IACTs
(OFF sample of Mrk 501) as well as from the Monte Carlo 
simulations are shown in Table 5. It is seen from Table 5 that after  applying the angular cut 
of $\Theta^2_0$ = 0.03 $\rm [deg^2]$ the cosmic ray background rejection is as high as 
$\simeq 200$. One can see also from Table 5 that Monte Carlo 
simulations reproduce quite well the measured contamination of the cosmic ray air 
showers after application of an angular cut. 

\subsection{Localization of Shower Core}

The position of the shower core at the observation level can be measured by 
the system of IACTs for a single individual event. Then the impact distances 
from the shower core to the system telescopes can be evaluated. The 
reconstruction algorithm is based on the orientation of the \v{C}erenkov light 
images in several telescopes which have been triggered. We use pure 
geometric reconstruction which does not relate to the image shape. The accuracy 
of the shower core reconstruction is limited by the errors in the 
determination of the image orientation. 
As discussed before, the change of the \v{C}erenkov light 
image topology leads to an increase of the error in the core position 
with increasing zenith angle. The accuracies of shower core reconstruction 
for the different primary energy and impact distance are summarized in Table 6. 
Observing a $\gamma$-ray source at zenith angles below 45 degrees and 
restricting the impact distances from the telescope to the shower core within 
200 m the average accuracy is $\leq 20$ m. 

The reconstructed impact distance is used for calculating the image shape 
parameters scaled on an impact distance and image amplitude. These 
parameters are applied for the cosmic ray rejection. For that,  
the accuracy of 20 m is quite sufficient because the shape of the \v{C}erenkov 
light images does not significantly change within 20 m. Furthermore, the value of 
the reconstructed impact distance is needed for evaluation of the shower 
energy. The calculations show that even with an accuracy of the shower 
core localization of around 20 m the energy resolution for 
$\gamma$-ray-induced air showers is better than 20\%.

\subsection{Measurement of Shower Energy}

The procedure of energy reconstruction for the $\gamma$-ray-induced air 
showers observed with a single imaging air \v{C}erenkov telescope appears to  
be quite complicated due to the lack of a direct measurements of the 
distance from the telescope to the shower core. Several indirect 
methods have been invented in order to estimate the impact distance using 
the centroid shift in the camera field of view  
as well as the image shape [19,20]. 

For the system of IACTs the measurement of the impact distance is 
straightforward and is not related to the image shape. That improves the 
accuracy in the energy reconstruction compared with a single telescope. 
The algorithms of the energy reconstruction for each individual event 
as well as for the spectrum evaluation with the system of IACTs have been 
discussed in [4,5,21]. This method of 
energy reconstruction was investigated by means of Monte Carlo 
simulations for the HEGRA system of IACTs. 

If one can measure the distance from the shower core to the telescope 
($r_i, i=1,N$, where $N$ is the number of triggered telescopes) the 
primary energy of air showers can be evaluated using the inverse function 
of the image size (amplitude) on shower energy and impact distance
\begin{equation}
E_i = F(S_i,r_i,\theta)
\end{equation}
where $S_i$ is the image size (total number of photoelectrons in the image) and 
$\theta$ is a zenith angle. 
The final energy estimate can be constructed by incorporating several images 
in a number of telescopes, $E_i, i=1,N$, as 
$E_0=\sum_i w_i E_i$, where $w_i$ is an energy an core distance dependent weight
($\sum w_i = 1$). Note that 
the energy resolution depends on the accuracy of determing the core 
distance and is strongly influenced by fluctuations in the image 
amplitude. 

In Table 7 we show the estimates of the energy resolution in the 
different primary energy ranges as well as for the simulations at different 
zenith angles. 
Note that in a case of {\it wobble} mode observations the images 
of high energy $\gamma$-ray-induced air showers $\rm E\geq 10 \, TeV$ are 
very often cut by the camera edge. This leads to a distortion in 
the impact distance reconstruction and consequently in the reconstruction 
of the shower energy. To remove this effect the large impact distances 
$r\geq 200$ m are usually excluded from the consideration. For 
inclined showers (at large zenith angles) the \v{C}erenkov light images are 
closer to the center of the field of view because these showers 
develop in atmosphere very far from the telescope and the corresponding 
images shrink to the camera center. Thus the problem of the 
camera edge appears to be less significant for inclined showers in the 
energy range at least up to 20 TeV. 
Data shown in Table 7 demonstrate that over the whole  
energy range for the $\gamma$-ray-induced air showers in observations at 
zenith angles up to 45 degrees the estimated energy resolution for the 
HEGRA system of IACTs is around $20$\%. 
 
\subsection{Rejection of CR background using the image shape}

Due to the difference in the nature of $\gamma$-ray- and proton/nuclei-induced 
air showers the corresponding \v{C}erenkov light images are also very different 
in shape [22]. In the standard second-moment approach these 
differences can be determined by using such image shape parameters as 
{\it Width, Length} etc. The $\gamma$-ray-induced air showers have on average, 
images of smaller angular size. The selection of the $\gamma$-rays can  
be done by applying the standard {\it parameters cuts}: {\it Width $\leq w_0, 
Length \leq l_0$} etc where {\it $w_0, l_0$} are the boundaries which limit 
the domain of the most of the $\gamma$-ray-induced showers ($\geq 50$\%).  

When the minimum amplitude for the pixel used for the image parametrization is fixed, the 
parameters of the image shape would increase with the total number of 
photoelectrons in the image. For the high energy air showers 
more pixels are involved in the image parametrization procedure that is why 
the angular size of the image increases with the shower energy. 
Application of fixed {\it image shape cuts} saves the most of the low 
energy $\gamma$-rays (which are close to the energy threshold) and reduces 
significantly the content of high energy $\gamma$-rays. For the 
$\gamma$-ray spectrum evaluation that is undesirable because it will 
decrease the statistics of the observed high energy $\gamma$-rays. To avoid 
this problem the energy dependent cuts have to be used. In addition, 
for a fixed primary energy of $\gamma$-ray-induced air showers the angular 
size of image also depends on the distance from the telescope to the 
shower core due to the decrease in image amplitude with impact distance. 
Thus for a single telescope using the restriction on the 
position of the image in a certain range from the camera center,  
one can keep only events at core distances 
$r\leq 120$ m. The radial dependence of the angular 
size of an image is very small in this particular range of impact distances. 
However, such a restriction significantly reduces the 
number of detected $\gamma$-rays.
For the system of IACTs both the energy and radial dependence of the image size 
can be accounted in detail, because the IACTs system measures the shower core 
position and consequently the energy for each individual event. 

The images from the $\gamma$-ray-induced air showers 
can be sorted into several bins on the measured distance from the shower 
core, $\Delta r_i, i=1,n$, and image size, $\Delta log(S_j), j=1,m$. Then the 
averaged image shape parameters are calculated for these particular bins, 
$<w>_{ij}, <l>_{ij}$. For an individual event the shower core position and 
impact distance from the shower core to the triggered telescopes can be 
reconstructed. 
Instead of the usual {\it Width } ($w$) of the image the {\it scaled Width} 
[17] ($\tilde w $)  is calculated for each telescope and after 
that the {\it mean scaled Width} parameter is defined for this event:
\begin{equation}
< \tilde w > = 1/N \sum_{k=1}^{N} w^k/<w>^k_{ij}
\end{equation} 
where $N$ is the number of triggered telescopes.
The rejection of cosmic ray background events is performed by applying a 
cut on the {\it mean scaled Width}, $< \tilde w > \leq < \tilde w_0 >$ . 
Measured distributions of the {\it mean scaled Width} for $\gamma$-ray-induced 
air showers observed at different zenith angles $\leq 10$, 30, 45 degrees are 
shown in Figure 6. The distributions have a peak at 1.0 and are very narrow  
because after the 
{\it scaling} procedure the RMS of the distributions is 
determined only by the pure image fluctuations and does not depend any more 
on the image amplitude and shower impact distance. The multifold cut on the 
image {\it Width} is replaced by single parameter {\it mean scaled 
Width} and works effectively against the cosmic ray background events. For a 
single telescope (i.e. one view of the shower) the probability, that 
cosmic ray shower gives an image beeing very similar to the $\gamma$-ray-induced 
shower, could be as high as 0.1. In observations by several 
telescopes (in different projections) this probability is reduced 
down to $10^{-2}$ (see Table 8). Note that in addition to {\it mean scaled 
Width cut} another shape parameters, such as {\it Length}, could be used. 
However, the simulations show that already after the {\it mean scaled Width cut} the 
cosmic ray background rejection is already very strong and an application of 
another shape cut does not improve the signal-to-noise ratio but simply reduces 
the amount of the $\gamma$-rays.
The acceptances of the cosmic ray showers after the application of a mean 
scaled Width cut can be compared for Monte Carlo simulations and observed 
cosmic ray events. The data shown in Table 8 demonstrates a good agreement 
between the simulations and data. 

There are two possible strategies for the choice of cuts. The first 
one is to apply the so-called strong cuts ($<\tilde w_0> = 1.0$) which dramatically 
suppress the CR background 
but at the same time also noticeably reduce the content of  
$\gamma$-rays (see Table 8). This approach works best when searching for the 
$\gamma$-ray signal from source candidates on a short time scale. For spectrum studies it is 
more effective to apply the loose cuts ($<\tilde w_0> = 1.4$) 
which save most of the observed $\gamma$-rays 
at high energies and do not show strong energy dependent efficiencies 
(opposite to the tight cuts). 
It is important to note that the efficiency of the cosmic ray rejection 
improves approximately by a factor of 2 for 3-telescope 
events compared to 2-fold coincidence events while the decrease of the $\gamma$-ray 
acceptance is only of $\simeq20$\%. Thus use of 3-fold events seems to be 
preferable from the point of view of the cosmic ray rejection using the image 
shape and 
orientation as well as for an accurate impact distance and energy reconstruction. 
However, in the specific case of the $\gamma$-ray fluxes characterized by very 
steep energy spectra the analysis of 2-fold coincidences could be applied  
in order to increase the statistics of $\gamma$-rays.  

\section{Sensitivity to $\gamma$-ray fluxes}

Using the calculated characteristics of the IACT system performance, 
one can estimate the sensitivity of the instrument in $\gamma$-ray observations.
For two complementary approaches of the cosmic ray background rejection 
based on the application of {\it tight} and {\it loose} cuts  
the acceptances of the $\gamma$-rays and cosmic rays are shown in Table 9.   
For the integral flux of $\gamma$-rays taken at the level of 
$J_\gamma(>1\,\rm TeV)=10^{-11} \rm \, cm^{-2}s^{-1}$ the expected 
number of detected $\gamma$-rays per one hour is about 28 and 56 for the 
{\it tight} and {\it loose} cuts, respectively. For the {\it tight} cuts the 
cosmic rays are highly suppressed and the expected cosmic ray rate per 
hour is around 3 particles. Thus, the stereoscopic observations of $\gamma$-ray 
sources of an integral flux 
$\rm J_\gamma(>1\, TeV)\geq 10^{-11} \, cm^{-2}s^{-1}$ 
with the HEGRA array of IACTs are essentially {\it background free} 
(rate of the detected $\gamma$-rays exceeds 
the rate of the background event by more than a factor of 10).  
The expected sensitivity 
of the 5 IACT system for the $\gamma$-ray flux stated above after application 
of the {\it loose} cuts is about  $5$ ``sigma-per-one-hour''. 
The 5$\sigma$ detection of $\gamma$-ray fluxes from point sources at 
the level of $J_\gamma(>1\,\rm  TeV)=10^{-12} \, cm^{-2}s^{-1}$ is 
possible in 20 hours of observations.  
Note, that after application of the {\it tight} cuts the amount of the 
detected cosmic electrons ($\simeq  0.4$ particle per hour) becomes comparable 
to the cosmic ray contamination and further suppression of the background 
is limited by the electron content. 

\section{Summary}

The HEGRA system of imaging air \v{C}erenkov telescopes is the first instrument 
operating in the stereoscopic observation mode. The HEGRA system of 5 telescope 
with relatively small mirror area of 8.5 $\rm m^2$ gives a rather low energy threshold 
of $\simeq 500$ GeV. The large camera field of view ($\simeq 4.3$ degree) allows to 
perform the observations of the point $\gamma$-ray sources in the wobble mode, 
taking at the same time the ON and OFF events, and increase the available observation 
time by a factor of two. The use of several images gives the angular resolution 
better than 0.1 degree and yields the cosmic ray rejection, using the shower 
orientation, up to 200 times. The geometrical reconstruction of the shower impact position 
with a good accuracy ($\leq 20$ m) improves the energy resolution and makes possible 
to apply the image shape cuts which are independent on a shower energy. The energy 
reconstruction procedure for the telescope system is straightforward and is not related 
to the image shape. The energy resolution of the HEGRA system in the dynamic energy range 
of 0.5 GeV - 30 TeV is better than 20 \%. In order to avoid the uncertainties in 
the evaluation of the energy spectrum of the detected $\gamma$-rays the effective 
collection area can be taken simply as the geometrical area around the center of the 
telescope system using the restriction on the reconstructed impact parameter (e.g., 
200 m). Simulateneos registration of a several \v{C}erenkov light images from the 
individual air shower makes possible to apply the correlated analysis of the image 
shape (mean scaled Width) and substantially improve the cosmic ray rejection up to 100. 
Note that the data analysis based on three concidence view appears to be preferable for 
the better angular, energy resolution as well as gives better cosmic ray rejection 
using the image shape. The three concidence events are optimum for the stereo imaging of 
the TeV $\gamma$-ray air showers. In the search mode the tight software cuts on shape 
and orientation show the better performance approaching the case of a background free 
detection of the $\gamma$-ray sources. However for the energy spectrum studies one 
can use the loose cuts providing the high $\gamma$-ray statistics.      
 
The HEGRA system of IACTs could be considered as a successful prototype 
for the future low energy (100 GeV) arrays such as HESS and VERITAS. In general the 
current experience of the HEGRA operation can be used for the design of the 
forthcoming arrays of IACTs.   

\section{Acknowledgements}

The support of the German Ministery for Research and Technology BMBF and of the Spanish 
Research Council CYCIT is gratefully acknowledged. We thank the Instituto de Astrophysica 
de Canarias for the use of the site and providing excellent working conditions. We 
gratefully acknowledge the technical support staff of Heidelberg, Kiel, Munich, and 
Yerevan. 

We thank the anonymous referee for suggesting the improvements in the manuscript.   

\pagebreak

\newpage 
\pagebreak

\begin{figure}[htbp]
\begin{center}
\epsfxsize= 16 cm
\epsffile[3 3 480 480]{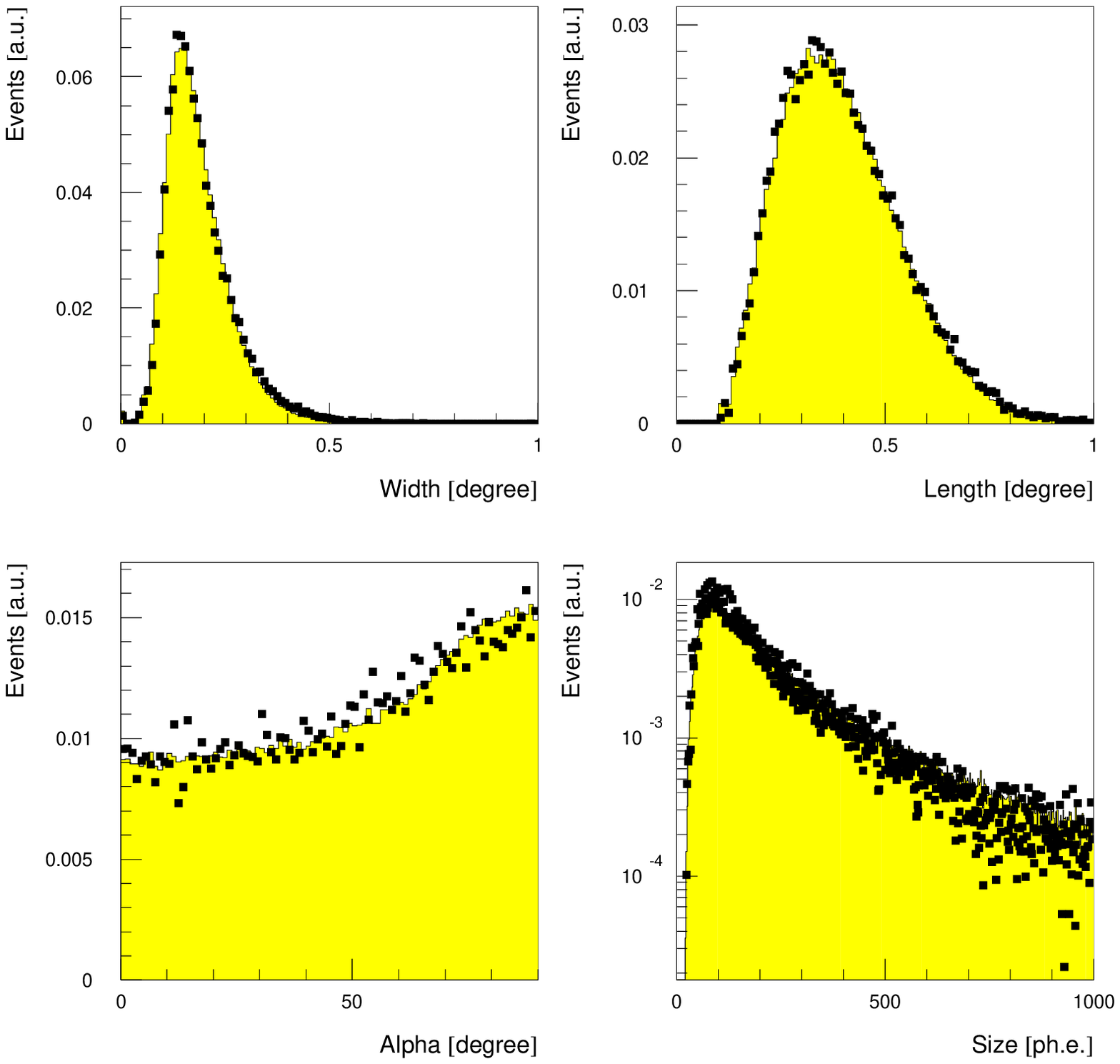}
\caption{\protect \small Distributions of image parameters of the recorded cosmic ray 
air showers and of Monte Carlo simulated cosmic ray air showers at the zenith angles 
up to 30 degrees. The hatched histograms correspond to the data, the points are for 
the Monte Carlo simulations.} 
\end{center}
\end{figure}

\pagebreak 

\begin{figure}[htbp]
\epsfxsize= 8 cm
\epsffile[2 21 292 292]{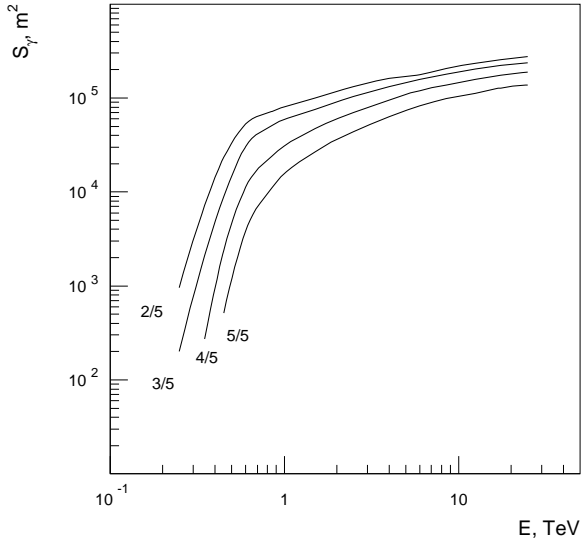}
\caption{\protect \small Collection areas of the $\gamma$-ray-induced air showers 
simulated at zenith for the complete array of 5 HEGRA IACTs. Data correspond to 
different trigger multiplicities N = 2, 3, 4 and 5 (N is a number of triggered 
telescope for an individual event). Local trigger for each telescope is taken as 
2nn/271$>$ 10 ph.e.}
\end{figure}

\begin{figure}[htbp]
\epsfxsize= 8 cm
\epsffile[3 7 300 300]{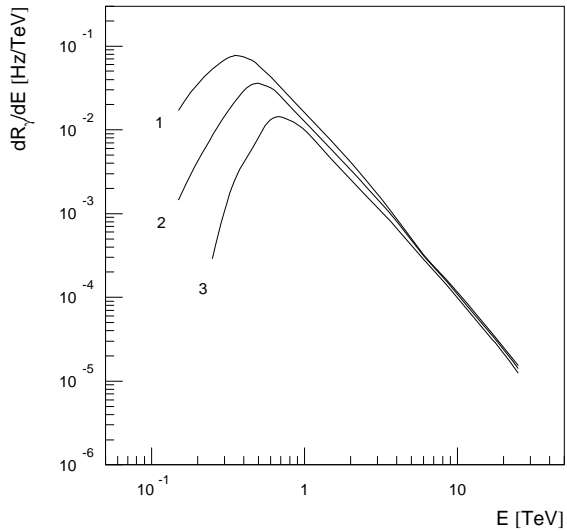}
\caption{\protect \small The differential detection rates of $\gamma$-ray-induced air 
showers simulated at $0^\circ$ zenith angle and detected by the system of 5 HEGRA 
IACTs operating with different local trigger condition: 2/5, $\rm 2nn/271>q_0$, 
$q_0$= 8 (1); 10 (2); 14 (3) ph. e. The assumed $\gamma$-ray spectrum is a power-law 
with the spectral index of 2.5 and with a flux normalization chosen according to Eqn. 
4.} 
\end{figure}

\pagebreak

\begin{figure}[htbp]
\vspace{8 cm}
\includegraphics{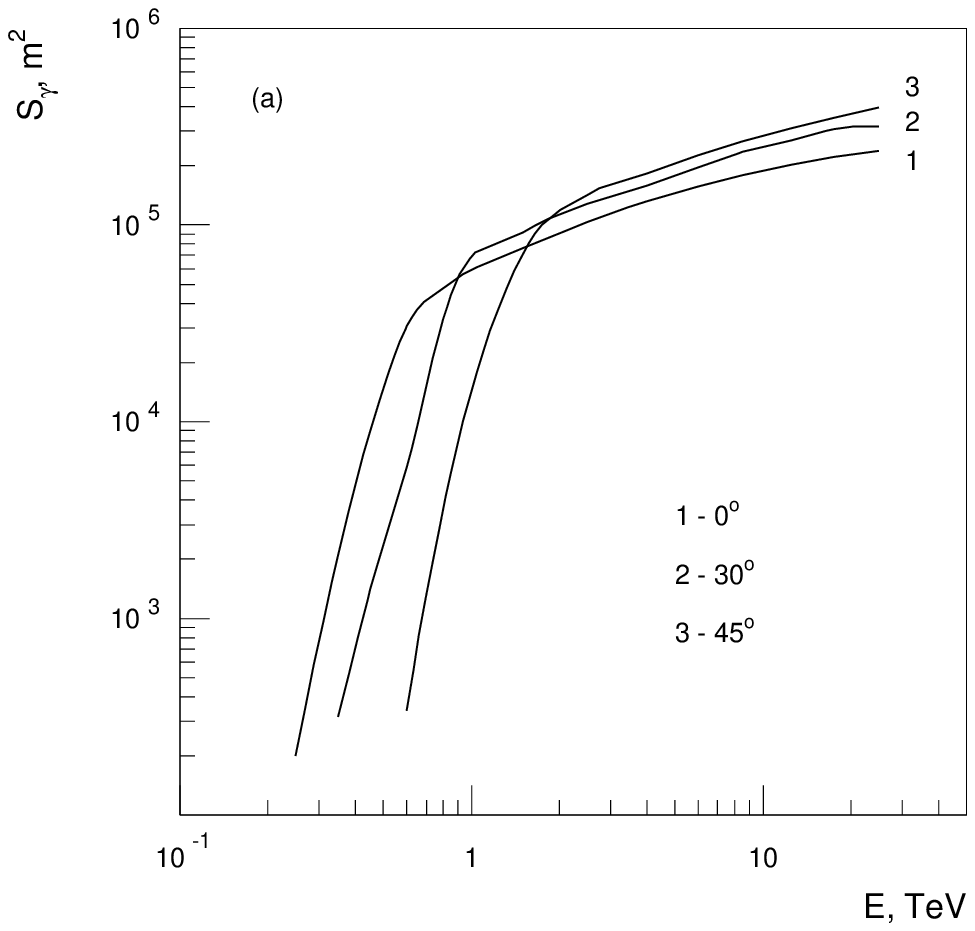}
\includegraphics{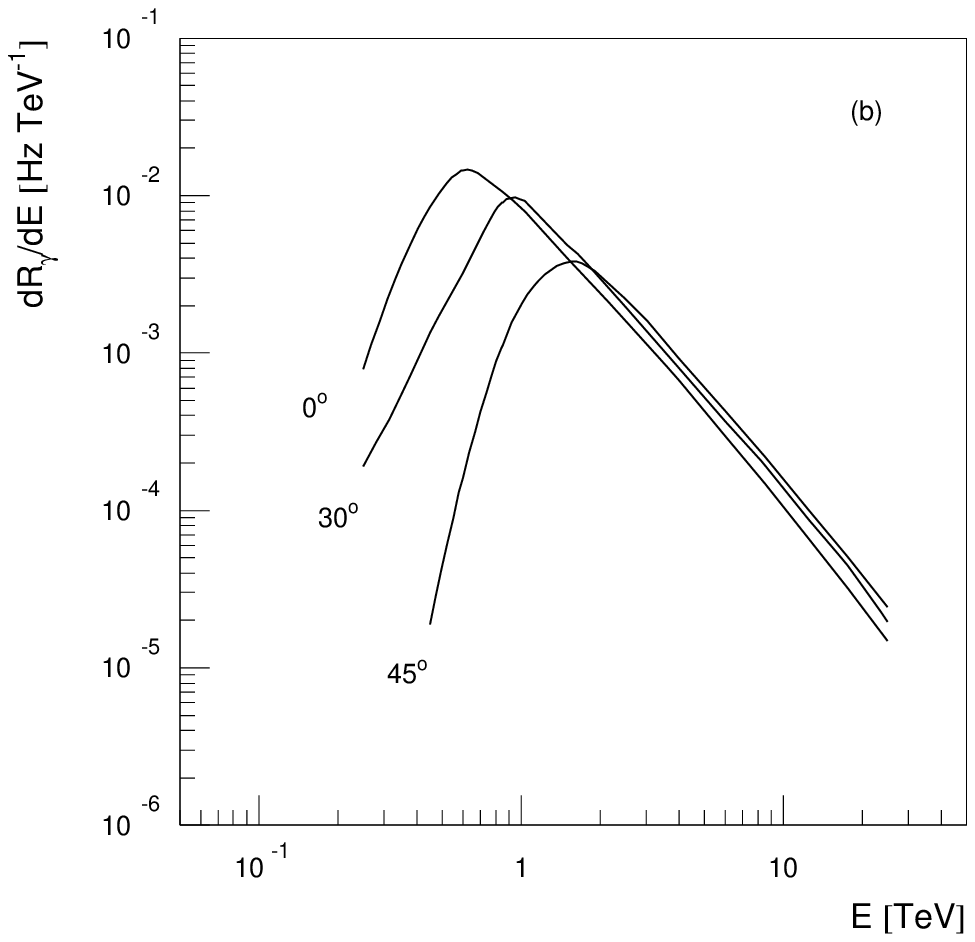}
\caption{\protect \small The collection areas (a) and differential detection rates (b) 
for the $\gamma$-ray-induced air showers simulated at zenith angles $0^\circ,\, 
30^\circ, \, 45^\circ$ and system trigger: 3/5, 2/271$>$10 ph.e. The assumed $\gamma$-
ray spectrum is a power-law with the spectral index of 2.5 and with a flux 
normalization chosen according to Eqn. 4.}
\end{figure}

\begin{figure}[htbp]
\epsfxsize= 9 cm
\epsffile[1 3 393 338]{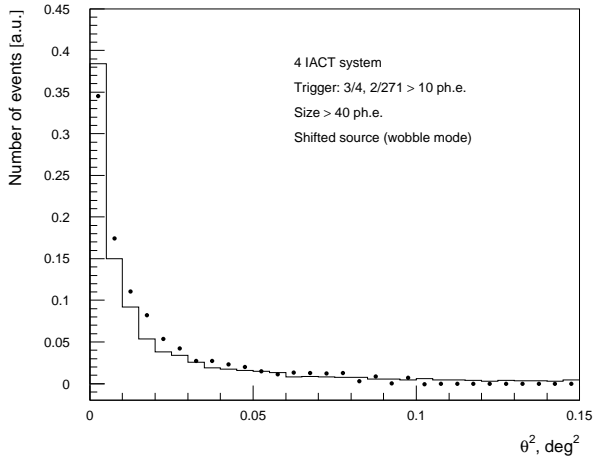}
\caption{\protect \small Distribution of reconstructed arrival directions of the 
$\gamma$-ray-induced air showers. $\Theta$ is an angular distance of the 
reconstructed position of the source in the camera focal plane from the true source 
position. The histogram corresponds to the Monte Carlo simulations. The distribution 
for $\gamma$-rays detected from Mrk 501 is indicated by black dots.}
\end{figure}

\pagebreak 

\begin{figure}[H]
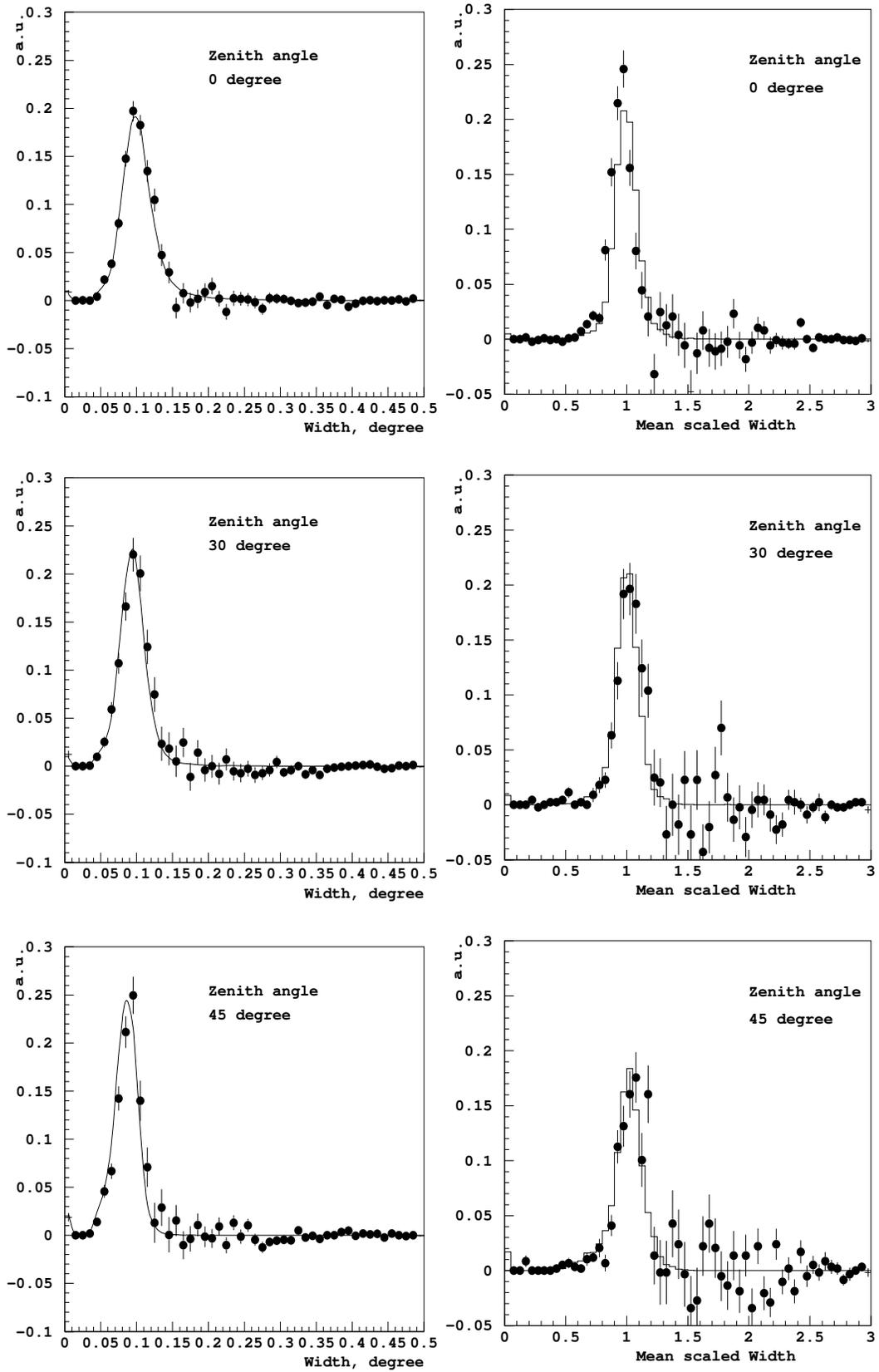

\begin{center}
\includegraphics[width=0.43\linewidth]{wd0-1.epsi}
\includegraphics[width=0.43\linewidth]{wd0-2.epsi}
\\\vspace{0.5cm}
\includegraphics[width=0.43\linewidth]{wd30-1.epsi}
\includegraphics[width=0.43\linewidth]{wd30-2.epsi}
\\\vspace{0.5cm}
\includegraphics[width=0.43\linewidth]{wd45-1.epsi}
\includegraphics[width=0.43\linewidth]{wd45-2.epsi}
\caption{\protect \small Distributions of {\it Width} (left column) 
and {\it mean scaled Width} (right column) parameter of the $\gamma$-ray-induced air 
showers detected by the HEGRA IACT array (dots) at different zenith angles $\sim$0, 
30, 45 degrees. Curves and histograms show the results of the Monte Carlo 
simulations.}
\end{center}
\end{figure}

\pagebreak 
\newpage

\begin{table}[htbp]
\caption{\protect \small 
The cosmic ray detection rates, $\rm R_{cr}$ [Hz], for the currently operating HEGRA 
system of 4 IACTs in observations at the zenith angles up to 30 degree. The data 
correspond to the different multi-telescope triggers: $N/4$ (N = 2, 3, 4) in the case 
of trigger $\rm 2nn/271>q_0$ for each individual telescope. The 
measured rates are taken from [11] (data are interpolated for the particular trigger 
thresholds).}
\label{rates_comp}
\bigskip
\begin{center} 
   \begin{tabular}{clllll} \hline \hline
 Trigger: & $q_0$ [ph.e.]= &   8  & 10  & 15  & 30 \\ \hline
  2/4     & (Data)         & 16.  & 9.6 & 5.5 & 2.4 \\
	  & (MC)           & 18.  & 10. & 5.6 & 2.0 \\ \hline       
  3/4     & (Data)         & 8.5  & 4.7 & 3.0 & 1.2 \\
	  & (MC)           & 9.0  & 5.1 & 2.7 & 0.9 \\ \hline
  4/4     & (Data)         & 3.8  & 1.8 & 1.3 & 0.5 \\
	  & (MC)           & 3.6  & 1.9 & 1.0 & 0.3 \\ \hline \hline
\end{tabular}
\end{center}
\end{table}

\begin{table}[htbp]
\caption{\protect \small The cosmic ray detection rates, $\rm R_{cr} [Hz]$, for the 
complete system of 5 HEGRA IACTs calculated at the zenith angles up to 30 degree for 
different multi-telescope triggers, $N/5$, and for two trigger thresholds $q_0 = 8, 
10$ ph.e.}
\vspace{0.3 cm} 
\begin{center}
\begin{tabular}{cllll} \hline \hline
$q_0$ ph.e. & 2/5  & 3/5  & 4/5 & 5/5 \\ \hline
    8       & 20.6 & 12.4 & 6.7 & 3.5 \\
   10       & 13.6 & 8.2  & 4.4 & 2.3 \\  \hline \hline
\end{tabular}  
\end{center}
\end{table}

\pagebreak

\begin{table}
\caption{\protect \small The detection rates of the $\gamma$-ray-induced air 
showers, $R_\gamma$ [photons/hour], for the complete HEGRA system of 5 IACTs for the 
different multi-telescope triggers. The local trigger for each individual telescope 
is set as $\rm 2nn/271>q_0, \, q_0$= 10 ph.e. The results correspond to a differential 
$\gamma$-ray spectrum in form of a power law with different spectrum index 
$\alpha_\gamma$.}
\vspace{3 mm}
\begin{center}
\begin{tabular}{cllll} \hline \hline
$\alpha_\gamma$ & 2/5 & 3/5 & 4/5 & 5/5 \\ \hline 
2.0 & 75.6 & 54.0 & 32.4 & 19.4 \\
2.5 & 97.2 & 61.2 & 34.2 & 17.3 \\
3.0 & 133.2& 79.2 & 36.0 & 17.6 \\ \hline \hline
\end{tabular}
\end{center}
\end{table}

\begin{table}[htbp]
\caption{\protect \small The Monte Carlo acceptance, $\kappa^{dir}_{\gamma}$, of the 
$\gamma$-ray-induced air showers simulated at different zenith angles, $\theta$, 
after application of the angular cut $\Theta^2<\Theta_0^2 \rm \, [deg^2]$. The 
calculations have been done for the system trigger of 2/5 and 3/5 and the local 
telescope trigger 2nn/271$>$10 ph.e. Additionally, the restriction on the distance 
of the shower core from the central telescope $\leq 200 \rm \, m$ was applied.}
\vspace{0.5 cm}
\begin{center}
\begin{tabular}{cclll}\hline \hline
Trigger: & $\Theta_0^2$ & $\theta=0^\circ$ & $30^\circ$ & $45^\circ$ \\ \hline
    & 0.03       & 0.72             & 0.68       & 0.58      \\ 
2/5 & 0.05       & 0.80             & 0.75       & 0.67      \\
    & 0.10       & 0.89             & 0.87       & 0.81      \\ \hline
    & 0.03       & 0.78             & 0.72       & 0.61      \\
3/5 & 0.05       & 0.85             & 0.78       & 0.72      \\ 
    & 0.10       & 0.92             & 0.89       & 0.84      \\ \hline \hline
\end{tabular}
\end{center}
\end{table}

\begin{table}[htbp]
\caption{\protect \small Acceptances of cosmic ray air showers, $\kappa^{dir}_{cr}$, 
after application of the angular cut, $\Theta^2<\Theta_0^2$. Data and Monte Carlo 
(MC) simulations correspond to zenith angles up to 30 degree, system trigger 3/5 and 
the local telescope trigger 2nn/271$>$10 ph.e.}
\vspace{1 mm}
\begin{center}
\begin{tabular}{rllllll}\hline \hline
 & $\Theta^2_0$ = 0.01   & 0.02   & 0.03   & 0.05   & 0.07   & 0.1 \\ \hline
$\kappa^{dir}_{cr}$ (Data) & 1.7$\cdot 10^{-3}$ & 3.4$\cdot 10^{-3}$ & 5.1$\cdot 10^{-3}$ & 8.6$\cdot 10^{-3}$ & 1.2$\cdot 10^{-2}$ & 1.7$\cdot 10^{-2}$ \\  
$\kappa^{dir}_{cr}$ (MC)   & 2.1$\cdot 10^{-3}$ & 4.1$\cdot 10^{-3}$ & 5.5$\cdot 10^{-3}$ & 9.0$\cdot 10^{-3}$ & 1.2$\cdot 10^{-2}$ & 1.8$\cdot 10^{-2}$ \\ \hline \hline
\end{tabular}
\end{center}
\end{table}

\begin{table}[t]
\caption{\protect \small Accuracy in localization of shower core of the 
$\gamma$-ray-induced air showers, $\delta r$ (one standard deviation of the 
Gaussian approximation), simulated at different zenith 
angles, $\theta$, and impact distances from the center of IACT array, $r$. 
The impact distance ranges are numbered as 0-50 m (1); 
50-100 m (2); 100-150 m (3); 150-200 m (4); 200-250 m (5).} 
\vspace{1 mm}
\begin{center}
\begin{tabular}{clccccc}\hline \hline
$\theta$ & E [TeV]        &(1)&(2)&(3)&(4)&(5) \\ \hline 
	   & 1-3          &  9  & 11 & 15 & 15 & 14 \\
$\sim 30^\circ$ & 3-5     &  5  &  6 & 11 & 17 & 15 \\
	   & 5-10         &  3  &  5 &  8 & 13 & 22 \\ \hline
	   & 1-3          & 20  & 22 & 27 & 25 & 20 \\
$45^\circ$ & 3-5          & 11  & 14 & 22 & 30 & 26 \\
	   & 5-10         &  6  & 10 & 15 & 27 & 34 \\ \hline \hline
\end{tabular}
\end{center}
\end{table}

\pagebreak 

\begin{table}
\caption{\protect \small Energy resolution, $\Delta E/E$, for $\gamma$-ray-induced 
air showers simulated at the zenith angles of 0, 30, 45 degrees. The 
maximum impact distance is 200 m. Data correspond to a system 
trigger of 3/5, and a local telescope trigger 2nn/271$>$10 ph.e.}
\vspace{0.5 cm}
\begin{center}
\begin{tabular}{cllllllll}\hline \hline
$\theta$   & $E_\gamma$ [TeV] = & 0.7-1 & 1-3  & 3-5  & 5-7  & 7-10 & 10-15 & 15-20 \\ \hline
$30^\circ$ &                    & 0.18  & 0.18 & 0.16 & 0.15 & 0.16 & 0.15  & 0.16  \\ 
$45^\circ$ &                    &       & 0.22 & 0.15 & 0.14 & 0.13 & 0.15  & 0.14  \\ \hline \hline
\end{tabular}
\end{center}
\end{table}

\begin{table}[htbp]
\caption{\protect \small The acceptance of the $\gamma$-rays, $\kappa_\gamma$, 
the cosmic ray background contamination, $\kappa_{cr}$, after application of a 
cut on {\it mean scaled Width}, $<\tilde w_0>$. 
The Monte Carlo simulations and data correspond to a zenith angle range up 
to 30 degree.} 
\vspace{3 mm}
\begin{center}
\begin{tabular}{rllllll} \hline \hline
$< \tilde w_0>$ = & 0.95  & 1.0   & 1.1   & 1.2   & 1.3   & 1.4     \\ \hline
$\kappa_\gamma$ (MC)  & 0.32  & 0.53  & 0.86  & 0.96  & 0.99  & 1.0  \\
$\kappa_{cr}$ (Data)  & 0.008 & 0.011 & 0.027 & 0.074 & 0.145 & 0.26 \\ 
$\kappa_{cr}$ (MC)    & 0.007 & 0.010 & 0.029 & 0.084 & 0.147 & 0.26 \\ \hline \hline
\end{tabular}
\end{center}
\end{table}

\pagebreak 

\begin{table}
\caption{\protect \small The expected rates of $\gamma$-ray-induced and 
cosmic ray induced air showers, simulated at zenith angles up to 30 degree, before 
and after application of the angular and shape cuts. The estimated signal-to-noise 
ratio for the $\gamma$-ray source with the intensity of $J_\gamma(>1\, TeV)=10^{-11} 
\, cm^{-2}s^{-1}$ (power-law spectrum index was taken as $\alpha_\gamma = 2.5$).
Data corresponds to the system trigger 3/5, the local telescope trigger 
2nn/271$>$10 ph.e. and 200 m maximum distance from the shower axis.}
\bigskip 
{\bf \it Tight cuts ($W_0<$1.0, $\Theta^2_0<0.05 \rm \, [deg^2]$):} 
\bigskip
\begin{center}
\begin{tabular}{lllllll}\hline \hline
$i$      & $R_i$, Hz & $\kappa^{dir}_i$ & $\kappa^{shape}_i$ & $\tilde {R_i}$, $hour^{-1}$  \\ \hline 
$\gamma$ & $1.7\cdot 10^{-2}$   & 0.86             & 0.53               & 28 \\ 
$CR$     & 7.8                  & 0.009            & 0.01               & 3  \\ \hline \hline
\end{tabular}  
\end{center}
\bigskip
{\bf \it Loose cuts ($W_0<$1.3, $\Theta^2_0<0.1 \rm \, [deg^2]$):}
\bigskip
\begin{center}
\begin{tabular}{lllllll}\hline \hline
$i$      & $R_i$, Hz & $\kappa^{dir}_i$ & $\kappa^{shape}_i$ & $\tilde {R_i}$, $hour^{-1}$  \\ \hline
$\gamma$ & $1.7\cdot 10^{-2}$    & 0.92             & 0.99               & 56 \\
$CR$     & 7.8                   & 0.018            & 0.15               & 76 \\ \hline \hline
\end{tabular}
\end{center}
\end{table}

\end{document}